\documentclass[aps,prl,twocolumn,superscriptaddress]{revtex4}
\usepackage{amsmath}
\usepackage[colorlinks]{hyperref}
\usepackage{amssymb}
\usepackage{color}
\usepackage{graphicx,amsmath}
\hypersetup{colorlinks,citecolor=red,linkcolor=blue,urlcolor=blue}

\begin{document}

\title{Comment on paper:
"Evidence for Dirac flat band superconductivity enabled by quantum
geometry" Nature {\bf 614}, 440 (2023)}
\author{V. R. Shaginyan}\email{vrshag@thd.pnpi.spb.ru} \affiliation{Petersburg
Nuclear Physics Institute of NRC "Kurchatov Institute", Gatchina,
188300, Russia}\affiliation{Clark Atlanta University, Atlanta, GA
30314, USA} \author{A. Z. Msezane}\affiliation{Clark Atlanta
University, Atlanta, GA 30314, USA}  \author{G. S.
Japaridze}\affiliation{Clark Atlanta University, Atlanta, GA 30314,
USA}
\begin{abstract}
\end{abstract}
\maketitle

The authors of Ref. \cite{1} consider the superconductivity in a
twisted bilayer graphene and explore the profound effect of
vanishingly small velocity in a superconducting Dirac flat band
system. They claim: "In a flat band superconductor, the charge
carriers' group velocity $v_F$ is extremely slow. Superconductivity
therein is particularly intriguing, being related to the
long-standing mysteries of high-temperature superconductors and
heavy-fermion systems". They continue: "In conclusion, twisted
bilayer graphene (tBLG) requires us to face the challenge of
ultra-strong coupling superconductivity in flat band Dirac systems,
where, naively, there can be no transport or superconductivity".
The authors "naively" suggest that the superconductivity based on
flat bands has not been thoroughly studied and analyzed. The
emergence of flat bands has been predicted about thirty years ago
which was accompanied by the astonishing claim that the
superconducting gap $\Delta\propto g$, with $g$ being the
superconducting coupling constant \cite{2,3,4,5,6,7,8}.
Simultaneously Fermi systems with flat band represent both the new
liquid and the new state of matter \cite{3,5,6}. It was also shown
that an absolutely flat band supports the superconducting state,
with the finite superconducting order parameter and the zero gap
\cite{5,6,8}. If the coupling constant $g$ is finite, then the
superconducting phase transition temperature $T_c$ becomes finite
and the flat band vanishes, since flat bands should be tuned with
the superconducting state. As a result, the flat band becomes
slightly tilted and rounded off at its end points
\cite{5,8,9,10,11}, as predicted \cite{10}. Measurements of the
Fermi velocity $v_F$ in magic-angle twisted bilayer graphene as a
function of $T_c$ revealed that $T_c \propto v_F\propto 1/N_s$,
where $N_s$ is the density of states at the Fermi level \cite{12}.
It was also shown that the high-$T_c$ compounds $\rm
Bi_2Sr_2CaCu_2O_{8+x}$ exhibit similar behavior. This observation
challenges theories of high-$T_c$ superconductivity, since $v_F$
normally correlates negatively with $T_c$: $T_c\propto 1/v_F\propto
N_s$, see \cite{9,11,12} and references therein. The topological
fermion condensation quantum phase transition, leading to flat
bands is considered and the special behavior of the strongly
correlated Fermi systems is elucidated, a feature not exhibited by
common Fermi liquids described by the Landau Fermi-liquid theory
\cite{5,6,11}. Topical reviews \cite{5,11} and monograph \cite{6}
combine theoretical explanations and experimental data on the
thermodynamic, transport and relaxation properties collected on
heavy fermion metals, high-$T_c$ superconductors and strongly
correlated Fermi systems.

In summary, we have demonstrated that an absolutely flat band
retains the superconducting state at $T_c\to0$ with $v_F\to0$. When
$T_c$ becomes finite, the flat band disappears, since it must be
modified by the superconducting state. The authors \cite{1} have to
take into account that at $T_c>0$ the flat band distorts, becoming
tilted, and $T_c\propto v_F$ \cite{9,11,12}. Thus, the statement
"the charge carriers' group velocity $v_ F$ is extremely slow" is
incorrect and leads the authors \cite{1} to the conceptional
misunderstanding, confusing the reader.

\end{document}